\documentstyle[amssymb,aps,prb]{revtex}

\begin{document}
\title{A SIMPLE ANALYTICAL MODEL OF VORTEX LATTICE MELTING IN 2D SUPERCONDUCTORS.}
\author{V.Zhuravlev, and T.Maniv}
\address{Chemistry Department, Technion-Israel Institute of Technology, Haifa 32000,\\
Israel.}
\date{\today{}}
\maketitle

\begin{abstract}
The melting of the Abrikosov vortex lattice in a 2D type-II superconductor
at high magnetic fields is studied analytically within the framework of the
phenomenological Ginzburg-Landau theory. It is shown that local phase
fluctuations in the superconducting order parameter , associated with low
energies sliding motions of Bragg chains along the principal
crystallographic axes of the vortex lattice , lead to a weak first order
'melting' transition at a certain temperature $T_{m}$ , well below the mean
field $T_{c\text{ }}$, where the shear modulus drops abruptly to a nonzero
value. The residual shear modulus above $T_{m}$ decreases asymptotically to
zero with increasing temperature. Despite the large phase fluctuations, the
average positions of Bragg chains at fimite temperature correspond to a
regular vortex lattice , slightly distorted with respect to the triangular
Abrikosov lattice. It is also shown that a genuine long range phase
coherence exists only at zero temperature; however, below the melting point
the vortex state is very close to the triangular Abrikosov lattice. A study
of the size dependence of the structure factor at finite temperature
indicates the existence of quasi-long range order with $S\left( 
\overrightarrow{G}\right) \sim N^{\sigma }$, and $1/2<\sigma <1$, where
superconducting crystallites of correlated Bragg chains grow only along
pinning chains. This finding may suggest a very efficient way of generating
pinning defects in quasi 2D superconductors. Our results for the melting
temperature and for the entropy jump agree with the state of the art Monte
Carlo simulations.

\end{abstract}

\section{Introduction}

Many potentially important superconductors , such as some of the High $T_{c}$
cuprates or the organic charge transfer salts $\kappa -(BEDT-TTF)_{2}X$ ( 
\cite{saito90}), are highly anisotropic compounds with nearly
two-dimensional electronic structure. These compounds are extremely type-II
superconductors with very small inplane coherence length. Consequently, the
Ginzburg critical region is relatively large and so one expects drastic
deviations from the predictions of the mean field theory for these materials
, due to strong thermal fluctuations in the superconducting order parameter.

In the mixed state at very low temperatures amplitude fluctuations are
suppressed , but phase fluctuations can lead to the melting of the vortex
lattice at a certain magnetic field $H_{m}\left( T\right) $ , $%
H_{c1}<H_{m}<H_{c2}$ \cite{friem96},\cite{sas98}. A soft shear , Goldstone
mode, which can be described by long wavelength phase fluctuations \cite
{moo89} is responsible for this remarkable melting phenomenon. \
Unfortunately rigorous analytical approaches to this problem have
encountered fundamental difficulties: large order high temperature
perturbation expansion with Borel-Pade approximants to the low temperature
behavior \cite{rug76}, \cite{bre90} has no indication of an ordered vortex
lattice even at zero temperature. The existing non-perturbative approaches
have not completely clarified the situation: Renormalization group studies 
\cite{moo96} , as well as Monte Carlo simulation \cite{oneil92} have
predicted no crystal vortex state in a pure 2D superconductor (SC) at finite
temperature, while the novel functional integral formalism suggested in \cite
{tes94},\cite{her94}, has led to some kind of a vortex liquid freezing
transition without breaking the $U(1)$ symmetry. Several Monte Carlo
simulations, have recently shown \cite{kat93}$^{-}$\cite{sas94} that in a 2D
, SC a true vortex lattice melting phase transition takes place at finite
temperature and that the transition is of the first order \cite{tes91}.

In this paper we present a simple model of the vortex lattice melting in 2D
extremely type-II superconductors. Our model is based on the observation
that at low temperature the main correction to the mean field free energy
arises from ''Bragg chain fluctuations'', namely fluctuations which preserve
long range periodic order along a principal crystallographic axis in the
vortex lattice. This simplification reduces our 2D problem to a 1D one ,
which can then be solved exactly.

Our calculations show that the Abrikosov triangular lattice is subject to
strong phase fluctuations. As a result of these fluctuations the sharp mean
field transition into the Abrikosov lattice state becomes a smooth
crossover. Since the strength of the phase dependent terms in the SC free
energy is relatively small ($\sim 2\%$ of the SC condensation energy) the
scale of the crossover temperature, $T_{cm}$ , is well below the mean field $%
T_{c}$. At temperatures higher than $T_{cm}$ the vortex lattice transforms
to an ensemble of strongly uncorrelated vortices, fluctuating independently
around equilibrium lattice positions. It is found that because of a
discontinuous (rotational) symmetry change in the mean positions of vortices
there is a weak first order transition , superimposed on the smooth
solid-liquid crossover, which is reflected in a small jump of the vortex
system entropy at a certain melting temperature $T_{m}\approx T_{cm}$.
Calculation of the structure factor shows that exact long range
translational order exists only at zero temperature, in agreement with
previous results \cite{moo96}, \cite{nik95}, \cite{sas95}. However, below
the melting temperature $T_{m}$, the vortex state is very close to the
triangular Abrikosov lattice. Our results for various thermodynamic
parameters agree well with numerical calculations in Ref.\cite{kat93}.

\section{The sliding 'Bragg-Chains' model}

Our starting point is the Ginzburg-Landau ( GL ) free energy functional 
\begin{equation}
F_{GL}=\int d^{2}r\left[ -\alpha \left| \psi \left( \overrightarrow{r}%
\right) \right| ^{2}+\frac{1}{2}\beta \left| \psi \left( \overrightarrow{r}%
\right) \right| ^{4}\right]  \label{pf1}
\end{equation}
with the order parameter $\psi \left( \overrightarrow{r}\right) $, defined
on the subspace of the lowest Landau level (LLL) \cite{eil67}. This
approximation is valid at sufficiently low temperatures or high magnetic
fields ( i.e. for $k_{B}T\ll \hbar \omega _{c}$ ), when thermal excitations
to higher Landau levels in the condensate of Cooper pairs can be neglected.
All possible configurations of the order parameter in this subspace can be
taken into account by considering the free energy 
\begin{equation}
F=-k_{B}T\ln Z  \label{pf2}
\end{equation}
with the partition function $Z$ defined by the functional integral: 
\begin{equation}
Z=\int D\psi D\psi ^{\star }\exp \left[ -F_{GL}/k_{B}T\right]  \label{pf3}
\end{equation}
In Eq.(\ref{pf1}) $\alpha =\alpha ^{\prime }\left( H_{c2}-H\right) $ and $%
\beta $ are phenomenological constants. The integral in Eq.(\ref{pf3}) is
performed over all non equivalent states.

An arbitrary wave function (in the symmetric gauge) from the LLL-subspace
can be written as a one dimensional integral 
\begin{equation}
\psi \left( x,y\right) =e^{ixy}\int dqc\left( q\right) \phi _{q}\left(
x,y\right)
\end{equation}
where $\phi _{q}\left( x,y\right) =\exp \left[ iqx-\left( y+\frac{q}{2}%
\right) ^{2}\right] $ is a Landau function with an orbital center located at 
$-q/2$ along the $y$-axis. Note that all spatial lengths are measured in the
units of the magnetic length $a_{H}=\left( c\hbar /eH\right) ^{1/2}$. A
system of $N$-vortices , with size $L_{x}=a_{x}\sqrt{N}$ along the $x$%
-direction , where $a_{x}$ is an arbitrary constant, is described by $N$
coefficients $c_{i}\equiv c\left( q_{i}\right) $, $q_{i}=\frac{2\pi }{L_{x}}%
i $ with $i=-N/2+1,...\,,N/2$.

It is well known that the minimal value of the GL free energy functional is
obtained when only $\sqrt{N}$ coefficients from the whole set of $N$
coefficients are different from zero ( i.e. $\ c_{n\sqrt{N}}\neq 0$ , for $%
n=-\sqrt{N}/2+1,...,\sqrt{N}/2$ ). At sufficiently low temperatures , when
amplitude fluctuations are suppressed (see later) , this minimum corresponds
to the minimum of the Abrikosov parameter:

\begin{equation}
\beta _{a}=\left( \frac{1}{V}\int d^{2}r\left| \psi \right| ^{4}\right)
/\left( \frac{1}{V}\int d^{2}r\left| \psi \right| ^{2}\right) ^{2}
\label{abrikos}
\end{equation}
where $V$ is the volume of the superconductor. From the definition of $\beta
_{a}$ it is seen that the absolute minimum, $\beta _{a}=1,$ is obtained for
a spatially uniform order parameter. Any deviation from $\left| \psi \right|
=const$ leads to an increase in $\beta _{a}$. Under the constraint of the
LLL subspace, however, $\left| \psi \right| \neq 0$ can not be a constant (
since $\left| \psi \right| =0$ at the vortex cores ), and the minimum $\beta
_{a}=\beta _{A}\simeq 1,159$ is obtained for the triangular Abrikosov
lattice , which is the closest configuration to the homogeneous one. Other
periodic lattices yield small positive deviations from $\beta _{A}$ , while
any departure of $\psi $ from the quasi uniform distribution of the vortex
lattice towards a localized structure leads to a drastic increase of the
free energy\cite{ftn1}.

Thus, we conclude that the main correction to mean field order parameter
arises from fluctuations of $c_{n}$ and $a_{x}$ , and take: $\ $

\begin{equation}
c\left( q\right) =\sum\limits_{n=-\sqrt{N}/2+1}^{\sqrt{N}/2}c_{n}\delta
\left( q-\frac{2\pi n}{a_{x}}\right)  \label{pf4}
\end{equation}

Note that regardless of the choice of $c_{n}$, $\psi \left( x,y\right) $ is
a periodic function of $x$\thinspace with a period $a_{x}$. Therefore, the
used representation of $c\left( q\right) $ allows the order parameter to
fluctuate only along the $y$-direction. Each coefficient, $c_{n}=\left|
c_{n}\right| e^{i\varphi _{n}}$ , describes a set of $\sqrt{N}$ Landau
orbital centers , periodically arranged within a certain chain along the $x$
-axis ('Bragg chain'). These 'Bragg chains' ( Fig. 1) are allowed to slide
independently along their common axis, where the phase $\varphi _{n}$,
determines the relative position $x_{n}=-\varphi _{n}/q_{n}$ of the $n$-th
chain. The ideal lattice states are obtained by selecting $c_{n}^{\left(
L\right) }=c_{0}\exp \left( i\gamma n^{2}\right) $, $0\leq \gamma \leq \pi
/2 $ \cite{man92}. Then for an arbitrary rhombic lattice the lattice
constant $a_{x}$ in units of magnetic length and the angle $\Theta /2$
between the principal crystallographic axes are expressed through $\gamma $
as $a_{x}^{2}=\pi /\sqrt{1-\left( \gamma /\pi \right) ^{2}}$, $\cos \Theta
=\gamma /\pi $.

The partition function (\ref{pf3}) can be therefore approximated by the
multiple functional integral: 
\begin{equation}
Z\approx \int \prod\limits_{n}dc_{n}dc_{n}^{\ast }\exp \left( -f_{GL}\right)
,
\end{equation}

where

\begin{equation}
f_{GL}\equiv \frac{F_{GL}}{k_{B}T\sqrt{N}}=-\overline{\alpha }\sum_{n}\left|
c_{n}\right| ^{2}+\frac{\overline{\beta }}{2}\sum_{n,s,t}\lambda
^{s^{2}+t^{2}}c_{n}^{\star }c_{n+s+t}^{\star }c_{n+s}c_{n+t}  \label{pf5}
\end{equation}
with $\overline{\alpha }=\frac{\alpha a_{x}}{\sqrt{2\pi }k_{B}T}$, $%
\overline{\beta }=\frac{\beta a_{x}}{\sqrt{4\pi }k_{B}T}$, $\lambda =\exp
\left( -\frac{\pi ^{2}}{a_{x}^{2}}\right) $. The inclusion of the factor $%
\sqrt{N}$ in the denominator takes into account the fact that each term in
the GL free energy functional , Eq.(\ref{pf1}) \ (written in the discrete
representation (\ref{pf4})) , corresponding to a certain Bragg chain in the
vortex lattice, is degenerate $\sqrt{N}$ times. This degeneracy reflects the
freezing of all $\sqrt{N}$ internal degrees of freedom within a chain. As
discussed above, by relaxing these degrees of freedom the free energy
functional develops very large, highly unprobable fluctuations. Therefore,
by dividing $F_{GL}$ with $\sqrt{N\text{ }}$ we are left with a single
nondegenerate term , which represents the dominant degree of freedom for
each chain.

Note that the functional Eq.(\ref{pf5}) is invariant under an arbitrary
linear shift of the phases $\varphi _{n}$: 
\begin{equation}
\varphi _{n}^{\prime }=\varphi _{n}+an+b  \label{pf5a}
\end{equation}
This symmetry follows from the symmetry of the GL free energy Eq.(\ref{pf1})
under the magnetic translation group \cite{bro68}.

The functional (\ref{pf5}) has a set of local minima with a single amplitude
for the entire set of lattice configurations:

\[
\left| c_{n}\right| ^{2}=\frac{\overline{\alpha }}{\overline{\beta }%
\overline{\beta }_{a}},\;\varphi _{n}=\gamma n^{2},\;f_{s}=-\frac{\overline{%
\alpha }^{2}}{2\overline{\beta }\overline{\beta }_{a}} 
\]
where 
\[
\overline{\beta }_{a}=\sum\limits_{sp}\exp \left[ -z\left(
s^{2}+p^{2}\right) \right] \cos \left( 2\gamma sp\right) 
\]
is a reduced Abrikosov parameter , and $z=\pi ^{2}/a_{x}^{2}$. These minima
can be immediately obtained if we use the symmetry of the functional $f_{GL}$
under the translation $n\rightarrow n+1$ . The Abrikosov parameter $\beta
_{a}$, defined by (\ref{abrikos}), is proportional to $\overline{\beta }_{a}$%
: $\beta _{a}=\sqrt{\pi }\overline{\beta }_{a}/a_{x}$. The triangular
lattice , with $\beta _{a}=\beta _{A}$ (\cite{klein63}), corresponds to $%
\gamma =\pi /2$, $a_{x}^{2}=2\pi /\sqrt{3}$. The mean field condensation
energy per unit vortex is $F_{GL}/N=-\frac{\pi \alpha ^{2}}{2\beta \beta _{A}%
}$.

Our main approximation at this point is based on the small value of the
parameter $\lambda \simeq e^{-\pi }$, which enables us to neglect in Eq.(\ref
{pf5}) all terms of the order higher than $\lambda ^{2}$ , i.e. to retain ,
in addition to the first order terms in $\lambda $, only the leading order
terms in the phase ( $\varphi _{n}$) dependent part of the free energy.
Thus, up to this order in $\lambda $%
\begin{equation}
f_{GL}=-\overline{\alpha }\sum_{n}\left| c_{n}\right| ^{2}+\frac{\overline{%
\beta }}{2}\sum_{n}\left[ \left| c_{n}\right| ^{4}+4\lambda \left|
c_{n}\right| ^{2}\left| c_{n+1}\right| ^{2}+4\lambda ^{2}\left|
c_{n-1}\right| \left| c_{n+1}\right| \left| c_{n}\right| ^{2}\cos \chi _{n}%
\right]  \label{pf6}
\end{equation}
The angles $\chi _{n}$ are linear combinations of the phases $\varphi _{n}$, 
\begin{equation}
\chi _{n}=-2\varphi _{n}+\varphi _{n-1}+\varphi _{n+1},  \label{pf7}
\end{equation}
which are clearly invariant under the transformation (\ref{pf5a}).

\section{Shear Motions : The Melting Mechanism}

The model presented in the previous section is based on our observation that
the low lying excitations of the Abrikosov vortex lattice are associated
with the sliding motions of the lattice Bragg chains along the principal
crystallographic axes. These excitations are closely related to the soft
shear modes discussed by Moore \cite{moo89} in connection with the vortex
lattice melting.

Let us consider this analogy in a greater detail. Following Ref.( \cite
{moo89} ) we first invoke a perturbative approach with respect to the
Abrikosov vortex lattice solution $c_{n}^{\left( L\right) }$ , by defining
the displacements $b_{n}$ through $c_{n}=c_{n}^{\left( L\right) }\left(
1+b_{n}\right) $ , and their normal mode (phonon) coordinates $u_{k}^{\pm }=%
\frac{1}{N^{1/4}}\sum\limits_{n}\left( b_{n}\pm b_{n}^{\star }\right)
e^{ikn} $, with $k=\frac{2\pi }{\sqrt{N}}l$, $l=-\sqrt{N}/2.....\sqrt{N}/2$
, and then expanding the free energy functional ( \ref{pf6}) to second order
in these coordinates. Omitting the details of calculations our result can be
written as: 
\begin{equation}
\delta f_{GL}=\frac{2\overline{\alpha }^{2}}{\overline{\beta }\overline{%
\beta }_{a}}\sum_{k\geq 0}\left[ P_{k}\left| u_{k}^{+}\right|
^{2}+Q_{k}\left| u_{k}^{-}\right| ^{2}\right]  \label{pf8}
\end{equation}
where in the long wavelength limit $k\rightarrow 0$ , $P_{k}\rightarrow 2$,
and $Q_{k}\rightarrow -\lambda ^{2}k^{4}\cos 2\gamma $. Thus, it can be
readily shown that the relative variance of the order parameter diverges,
i.e. : $\left( \left\langle \psi ^{2}\right\rangle -\left\langle \psi
\right\rangle ^{2}\right) /\left\langle \psi \right\rangle ^{2}\sim \int 
\frac{dk}{k^{4}}$. Therefore, we can conclude that although the lattice
states with $\pi /4<\gamma <\pi /2$ are thermodynamically stable ($\delta
f_{GL}>0$), the soft modes $Q_{k}$ , $k\rightarrow 0$ , are responsible for
infinite fluctuations of the order parameter. This result is similar to that
obtained by Moore \cite{moo89} for a 2D system of fluctuations. The
divergence, which arises because of the perturbative nature of the above
calculation, is stronger in our 1D model. However, within the
nonperturbative method developed below this divergence is removed.

It should be noted that the soft mode described above is associated with the
long wavelength component of the phase fluctuations in Eq.(\ref{pf6}); this
can be seen by neglecting amplitude fluctuations, defining fluctuating
phases: 
\[
\varphi _{f}\left( n\right) \equiv \varphi _{n}-\varphi _{n}^{\left(
L\right) }=\varphi _{n}-\frac{1}{2}\pi n^{2} 
\]
and taking the continuous limit (see Eq.(\ref{pf7})), i.e. $\chi
_{n}\rightarrow \pi +\partial ^{2}\varphi _{f}/\partial y^{2}$ , so that the
relevant part in the free energy functional (\ref{pf6}) can be written as:

\begin{equation}
\delta f_{ph}=K_{A}\int \cos \left( \pi +\partial ^{2}\varphi _{f}/\partial
y^{2}\right) dy\approx \frac{1}{2}K_{A}\int \left( \partial ^{2}\varphi
_{f}/\partial y^{2}\right) ^{2}dy
\end{equation}
where $K_{A}\approx \lambda ^{2}\frac{\overline{\alpha }^{2}}{2\overline{%
\beta }}$. It is instructive to compare this expression with that derived in
(\cite{moo89}) for the effective Hamiltonian associated with a smoothly
varying phase $\theta \left( x,y\right) $, namely: $H_{ph}=\frac{1}{2}%
c_{66}\int d^{2}r\left( \nabla ^{2}\theta \right) ^{2}$, where $c_{66}$ is
an isotropic shear modulus of the vortex lattice, which is given
approximately by $\frac{1}{2}\frac{\overline{\alpha }^{2}}{\overline{\beta }}
$.

The agreement between the two approaches is , however, incomplete , not only
because of the one dimensional nature of our model (in contrast to the 2D
analysis of Ref(\cite{moo89})), but also because of the significant
difference in the 'stiffness' parameters $K_{A}$, and $c_{66}$ , namely $%
K_{A}/c_{66}\approx \lambda ^{2}\sim 10^{-2}$. \ The reason for the
disagreement can be understood within our approach by considering shear
motions along families of Bragg chains with Miller indices higher than of
the principal ones. For these families the values of $a_{x}$ are relatively
large , and the corresponding values of $\lambda ^{2}$ are not small
compared to unity. In the limit of very large Miller indices, $%
a_{x}\rightarrow \infty $ , and $\lambda ^{2}\rightarrow 1$, so that the
corresponding stiffness parameter approaches $c_{66}$, and becomes
independent of the chain orientation, as in Moore's theory.

Thus, in contrast to the isotropic shear model used in (\cite{moo89}) , the
appearance of the small parameter $\lambda ^{2}$ in front of the leading
phase dependent terms of the free energy functional (\ref{pf6}) implies that
the shear motions along the two principal crystallographic axes cost a small
fraction of the condensation energy , and so lead to significant distortions
of the vortex lattice along these particular directions at very low
temperatures ( i.e. with respect to the mean field $T_{c})$.

Considering this low temperatures regime, the calculation of the partition
function $Z$ can be simplified considerably since amplitude fluctuations can
be neglected. The functional integrals in Eq.(\ref{pf3}) over the order
parameter $\left\{ \psi ,\psi ^{\star }\right\} $ should be replaced by
integrals over the new variables $\left\{ c_{n},c_{n}^{\star }\right\}
\equiv \left\{ \left| c_{n}\right| ,\varphi _{n}\right\} $.

For an arbitrarily large system due to the invariance of free energy under
the magnetic translation group, the transformation Eq.(\ref{pf7}) is
degenerate, and the inverse transformation, $\varphi _{n}=\varphi _{n}\left(
\left\{ \chi _{m}\right\} \right) $, is not unique. It is determined up to a
linear function of $n$. To determine the phases $\varphi _{n}$ uniquely , we
have to impose additional conditions, determining the arbitrary constant $a$
and $b$ in Eq.(\ref{pf5a}). These conditions are equivalent to boundary
conditions of Eq.(\ref{pf7}). Linear boundary conditions lead to linear
dependence of $a$ and $b$ on $\chi _{m}$. Since the determinant of any such
transformation does not depend on the variables, the partition functions for
various $a$ and $b$ differ by a constant factor. Therefore, instead of $%
\left\{ \varphi _{n}\right\} $ one can integrate over $\left\{ \chi
_{n}\right\} $ with the same free energy functional. The GL free energy has
different values when the phases $\left\{ \chi _{n}\right\} $ lie within the
interval $[0,\pi ]$. To satisfy this condition and to exclude double
counting of fluctuation we integrate over interval $\chi _{n}\in [0,\pi ]$
and allow the phases $\varphi _{n}$ in Eq.(\ref{pf7}) to have arbitrary
values.

Omitting unimportant constant factor we obtain after integration over angle
variables that the partition function can be written as: 
\begin{equation}
Z=Z_{v}^{\sqrt{N}}\propto \int\limits_{0}^{\infty }\prod\limits_{n}\left|
c_{n}\right| d\left| c_{n}\right| e^{-f_{s}},  \label{pf9a}
\end{equation}
where 
\begin{equation}
f_{s}=\sum_{n}\left\{ -\overline{\alpha }\left| c_{n}\right| ^{2}+\frac{%
\overline{\beta }}{2}\left( \left| c_{n}\right| ^{4}+4\lambda \left|
c_{n}\right| ^{2}\left| c_{n+1}\right| ^{2}\right) -\frac{1}{\pi }\ln
I_{0}\left( 2\overline{\beta }\lambda ^{2}\pi \left| c_{n-1}\right| \left|
c_{n+1}\right| \left| c_{n}\right| ^{2}\right) \right\}  \label{pf9}
\end{equation}
and $I_{k}\left( x\right) $ is the modified Bessel function of the order $k$.

Neglecting amplitude fluctuations, the integrals in Eq.(\ref{pf9a}) can be
performed by the stationary phase approximation. Since the last term in Eq.(%
\ref{pf9}) is of the order $\lambda ^{2}$ or smaller, the approximate
solution to the stationary point equations, $\frac{\partial f_{s}}{\partial
\left| c_{n}\right| }=0$, can be simply obtained by using the translational
symmetry of the free energy functional $f_{s}$. It is similar to the mean
field solution, $\left| c_{n}\right| ^{2}=\frac{\overline{\alpha }}{%
\overline{\beta }\overline{\beta }_{fl}}$ with the generalized Abrikosov
parameter 
\begin{equation}
\overline{\beta }_{fl}\simeq 1+4\lambda -4\lambda ^{2}\frac{I_{1}\left( \tau
\right) }{I_{0}\left( \tau \right) }
\end{equation}
where 
\[
\tau =\frac{4\lambda ^{2}}{\left( 1+4\lambda \right) ^{2}}\frac{\pi 
\overline{\alpha }^{2}}{2\overline{\beta }}\equiv \frac{T_{cm}}{T} 
\]

The temperature $T_{cm}\left( a_{x}\right) $ determines a smooth crossover
from the mean field lattice state with $\gamma =\pi /2$, $\overline{\beta }%
_{fl}=\overline{\beta }_{l}\equiv 1+4\lambda -4\lambda ^{2}$, to a new state
corresponding to $\overline{\beta }_{fl}=\overline{\beta }_{m}\equiv
1+4\lambda $, where the phase dependent terms in the free energy are
completely destroyed by fluctuations. Note that the energy difference
between these states is of the order of the small parameter $\lambda ^{2}$.

In the zero temperature limit, $T\ll T_{cm}\left( a_{x}\right) $, the
parameter $\beta _{fl}\simeq \frac{\sqrt{\pi }}{a_{x}}\left( 1+4\lambda
-4\lambda ^{2}\right) $ has minimal values at $a_{x}^{2}=\frac{2\pi }{\sqrt{3%
}}$ , and $a_{x^{\prime }}^{2}=2\sqrt{3}\pi $ (Fig. 2) , depending on the
choice of the 'Bragg chains' family (i.e. along the $x$ or $x^{^{\prime }}$
axis in Fig. 1). Both of the minima describe a triangular Abrikosov lattice
with $\beta _{fl}=\beta _{A}\simeq 1.1596$. Both directions can be selected
in three equivalent ways in the Abrikosov lattice. All equivalent
configurations can be obtained from the invariance of the mean field
Abrikosov parameter $\beta _{a}=\sqrt{\frac{z}{\pi }}\sum\limits_{sp}\exp
\left( -z\left( s^{2}+p^{2}\right) \right) \cos \left( 2\gamma sp\right) $,
where $z=\pi ^{2}/a_{x}^{2}$, under the transformations $z^{\prime }=\frac{%
\pi ^{2}z}{z^{2}+\gamma ^{2}}$, $\gamma ^{\prime }=\frac{\pi ^{2}\gamma }{%
z^{2}+\gamma ^{2}}$ and $z^{\prime }=z$, $\gamma ^{\prime }=-\gamma $ or $%
\gamma ^{\prime }=\gamma +\pi n$ with an arbitrary integer $n$. \ 

The doubly degenerate , equilibrium state at $T=0$, just described, is
stabilized by the competition between two types of interactions among
parallel chains: The repulsive interaction between any two neighboring
chains, which is linear in the coupling parameter $\lambda $ , and the
attractive 3-body, phase dependent interaction ( i.e. involving any three
neighboring chains) , which is quadratic in $\lambda $ (see Eq.(\ref{pf6})).
At finite, low temperatures, i.e. when $T\sim T_{cm}$ , the shear
fluctuations destroy the phase coherence among parallel Bragg chains, thus
diminishing the small attractive interaction , and raising the total free
energy. The relatively large, repulsive interaction is affected only at
higher temperatures.

The interchain coupling parameter $\lambda $ , depends on the lattice
parameter $a_{x}$, through $\lambda =e^{-\pi ^{2}/a_{x}^{2}}$ . Since $%
a_{x^{^{\prime }}}>a_{x}$ (Fig. 1), the chains along $x^{\prime }$ are
closer to each other than those along $x$ , and so $\lambda \left(
a_{x^{\prime }}\right) >\lambda \left( a_{x}\right) $. Consequently, at low
temperatures , $T\lesssim T_{cm}\left( a_{x}\right) $ , when the attractive
3-body interaction diminishes with increasing temperature, the first state ($%
a_{x^{^{\prime }}}$) is more stable than the second one ($a_{x}$), since its
free energy increases more slowly with increasing temperature than that of
the second one (Fig. 3). At higher temperatures $T\gtrsim T_{cm}\left(
a_{x^{\prime }}\right) $, when the repulsive interchain couplings determine
the temperature dependence, the tendency is reversed and the free energy of
the first state ($a_{x^{^{\prime }}}$) increases faster with increasing
temperature than that of the second one ($a_{x}$). Thus, there is an
intersection point $T_{cm}\left( a_{x}\right) \lesssim T_{m}\lesssim
T_{cm}\left( a_{x^{\prime }}\right) $, at which the free energies of the
these states are equal, but the corresponding entropies are a little
different. Therefore, we conclude that at $T=T_{m}$ , there is a weak first
order transition characterized by a small jump of the lattice entropy.
Defining the parameter $t\equiv -\frac{2\pi \alpha ^{2}}{\beta k_{B}T}$ ( 
\cite{kat93}), the position of the crossing point corresponds to $%
t=t_{m}\simeq -16$ , and the jump in the entropy ( $S\equiv -T\frac{\partial
F}{\partial T}$ ) is $\Delta S\simeq 7.5\,10^{-3}F_{MF}/T$ . The values of $%
t_{m}$ and $\Delta S$ agree pretty well with the Monte-Carlo simulations 
\cite{kat93}.

The physical nature of this transition can be illuminated by considering the
shear modulus $\mu $. The vanishing of the shear modulus in atomic crystals
is usually regarded as a definition of the crystal melting point. In our
case $\mu $ can be calculated by transforming $c_{n}^{\prime }=e^{i\eta
n^{2}}c_{n}$ \cite{sas94} and taking the limit:

\[
\mu =\left( \frac{\partial ^{2}F_{GL}}{\partial \eta ^{2}}\right) _{\eta
\rightarrow 0} 
\]

Note that the considered transformation shifts the phases $\chi _{n}$ by $%
-2\eta $. Therefore, the shear modulus is proportional to the phase factor
in the free energy, i.e. $\mu \propto \left\langle \cos \chi
_{n}\right\rangle $. Normalized by the mean field value, $\mu _{MF}$, where $%
\cos \chi _{n}=-1$, it is reduced to 
\begin{equation}
\frac{\mu }{\mu _{MF}}=\frac{I_{1}\left( \tau \right) }{I_{0}\left( \tau
\right) }
\end{equation}

The dependence of the shear modulus on the parameter $t$ is plotted in
Fig.(4). At the transition point, $t=t_{m}$, the value of the parameter $%
a_{x}$ , corresponding to the minimum free energy, changes abruptly and the
shear modulus jumps from $\mu _{1}$ to $\mu _{2}$. It should be stressed
that $\mu _{2}\neq 0$. The residual shear energy on the high temperature
side of the transition point reflects an incomplete melting at $t=t_{m\text{ 
}}$. The 'liquid' state on this side of the transition point retains some
degree of phase coherence between different chains, which continues to
decrease gradually to zero with increasing temperature, reaching the
complete liquid state only asymptotically. This behavior seems to be due to
the persistence of long range periodic order along the chains axis in our
model at any temperature.

Interesting structural information on the 'quasi liquid' states described
above can be obtained from the calculation of the average values 
\[
\left\langle \chi _{n}^{k}\right\rangle =\frac{1}{\pi I_{0}\left( \tau
\right) }\int_{0}^{\pi }d\chi _{n}\chi _{n}^{k}\exp \left( -\tau \cos \chi
_{n}\right) 
\]

with $k=1,2$ . In the low and high temperature limits 
\begin{eqnarray}
\left\langle \chi _{n}\right\rangle &=&\pi -\sqrt{\frac{2}{\pi \tau }}%
\,\,\,\,\,\,\,\,\,\,\,\,\,\,\left\langle \chi _{n}^{2}\right\rangle =\pi
^{2}-\left( \frac{8\pi }{\tau }\right) ^{1/2}+\frac{1}{\tau }%
\,\,\,\,\,\,\,\,\,\,for\,\,\,\tau \gg 1  \label{pf14} \\
\left\langle \chi _{n}\right\rangle &=&\pi /2+2\tau /\pi
\,\,\,\,\,\,\,\,\,\,\,\,\,\,\,\left\langle \chi _{n}^{2}\right\rangle =\pi
^{2}/3+2\tau
\,\,\,\,\,\,\,\,\,\,\,\,\,\,\,\,\,\,\,\,\,\,\,\,\,\,\,\,\,\,\,\,\,\,\,\,%
\,for\,\,\,\tau \ll 1  \nonumber
\end{eqnarray}
The square root of the relative variance $\sigma =\sqrt{\left\langle \chi
_{n}^{2}\right\rangle -\left\langle \chi _{n}\right\rangle ^{2}}%
/\left\langle \chi _{n}\right\rangle $, is found to be $\sigma \simeq \frac{%
\pi -2}{\pi \tau }\ll 1$ in the low temperatures regime and $\sigma \simeq 1/%
\sqrt{3}$ in the high temperatures one. These results show that with the
temperature increase the fluctuations destroy the phase correlation between
chains so that the SC state transforms from a frozen Abrikosov lattice at
zero temperature to a new, \ 'liquid' \ state with strong vortex
fluctuations. However, in contrast to the usual liquid state , here we find
that the average vortex positions form a regular lattice with $\gamma =\pi
/4 $ and $\pi ^{2}/a_{x}^{2}\simeq 2.97$, where $a_{x}$ corresponds to the
minimum free energy. For this lattice the angle, $\Theta /2$ , between the
principal crystallographic axes corresponds to$\ \Theta \simeq 75^{\circ }$.

The first order 'melting' point at $t=t_{m}$ thus corresponds to a
discontinuous (rotational) symmetry breaking in this lattice of average
vortex positions, from $\Theta \simeq 60^{\circ }$ (Abrikosov lattice) on
the low temperature side , to $\Theta \simeq 75^{\circ }$ on the high
temperature one.

\section{Bragg-Chains Pinning and the absence of Long Range Order}

An intriguing issue in the theory of the vortex lattice melting concerns the
existence or the absence of long range phase coherence in the SC mixed
state. In this section we address the problem of long range order (LRO)\ and
the related topic of vortex lines pinning , as they appear in our model.

As discussed in Sec.(III), the fluctuating phases $\varphi _{n}$ can not be
uniquely determined from $\chi _{n}$; they depend on the choice of the
boundary conditions for Eq.(\ref{pf7}). Since the general solution of the
homogeneous equation ($\chi _{n}=0$) is a linear function of $n$ (Eq.(\ref
{pf5a}), two constants ($a$ and $b$) are required to uniquely determine $%
\varphi _{n}$ . A possible choice is to take $a=b=0$ , which corresponds to
the selection $\varphi _{0}=0$ , and $\varphi _{-1}=\varphi _{1}$. The
physical meaning of the first condition is that the chain , labeled $n=0$ (
i.e. located vertically at $y=0$) , is pinned to a fixed 'horizontal' (i.e.
along the $x$-axis) position. The second condition has a clear physical
meaning in the long wavelength limit , namely that the horizontal
displacement $u_{x}=\left( \partial \theta /\partial y\right) $ of the
vortex lines vanishes at the pinning site $y=0$ .

The solution of Eq.(\ref{pf7}) which satisfies this particular pinning
condition is ( for $n>0$) : 
\begin{equation}
\varphi _{\pm n}=\sum_{l=0}^{n}\left( n-l\right) \chi _{\pm l}+\frac{n}{2}%
\chi _{0}  \label{pf15}
\end{equation}

This transformation enables us to calculate any correlation function of
phase factors; in particular the pair correlation function:

\[
\langle e^{i(\varphi _{n^{\prime }}-\varphi _{n})}\rangle =\Pi _{\nu
}\int_{0}^{\pi }d\chi _{\nu }e^{-\tau \cos \chi _{\nu }}e^{i(\varphi
_{n^{\prime }}-\varphi _{n})}/\Pi _{\nu }\int_{0}^{\pi }d\chi _{\nu
}e^{-\tau \cos \chi _{\nu }} 
\]
can be readily evaluated by using Eq.(\ref{pf15}) to yield: 
\begin{equation}
\langle e^{i(\varphi _{n^{\prime }}-\varphi _{n})}\rangle =\frac{\Pi _{\nu
}^{n}I_{\upsilon }\left( -\tau \right) }{I_{0}\left( -\tau \right) ^{n+1}}
\end{equation}
where the function $\upsilon \equiv $ $\upsilon \left( \nu ;n,n^{\prime
}\right) $ $=\upsilon \left( \nu ;n^{\prime },n\right) $ is defined ( for $%
n>n^{\prime }$) \ by : $\frac{1}{2}\left| n-n^{\prime }\right| ,\;$at $\;\nu
=0$ ; $\;\left| n-n^{\prime }\right| ,\;$for$\;0<\nu \leq n^{\prime }$ ; $%
\left| n-\nu \right| ,$ for $\ n^{\prime }<\nu \leq n$ ; and $0,\ $for $\
\nu >n$. \ $\ $

In the high temperature limit $\tau \ll 1$ , far above the melting point,
the small argument expansion of the modified Bessel function $I_{\upsilon
}\left( -\tau \right) $ yields: 
\begin{equation}
\langle e^{i(\varphi _{n^{\prime }}-\varphi _{n})}\rangle \propto \tau ^{%
\frac{1}{2}\left( n^{2}-n^{\prime 2}\right) }\rightarrow \delta
_{n,n^{\prime }}
\end{equation}
meaning no phase correlation at all.

In the low temperature limit $\tau \gg 1$, the asymptotic expansion of $%
I_{\upsilon }$ $\left( -\tau \right) $ \ leads ( for any $n^{\prime }\gtrsim
n\gg 1)$ to the expression:

\begin{equation}
\langle e^{i(\varphi _{n^{\prime }}-\varphi _{n})}\rangle \approx e^{i\pi
(n^{\prime }-n^{2})/2}\exp \left[ -\frac{\overline{n}}{2\tau }(\Delta n)^{2}%
\right]  \label{pf16}
\end{equation}
where $\Delta n=n^{\prime }-n$, and $\overline{n}=n^{\prime }/3+2n/3-1/2$.
Eq.(\ref{pf16}) is identical to a second order cumulant expansion with $%
\left\langle \varphi _{n}\right\rangle =\frac{1}{2}\left\langle \chi
_{n}\right\rangle n^{2}\approx \frac{1}{2}\pi n^{2}$.

This result shows that a genuine long range phase correlation exists only at
zero temperature; it also shows that a cluster of highly correlated chains
can grow only along a pinned chain, since the phase fluctuations diverge
with the distance from the pinning chain (see Eq.(\ref{pf15})) . Note that
the position of such a chain is arbitrary since there is no energy cost to
pinning in the GL theory used. \ In real samples the translational symmetry
is broken by impurities , crystal defects , and the termination of the
lattice at the sample surface, which can pin chains of orbital centers to
fixed positions. A single pinning center located near a given chain may pin
the entire chain due to the chain rigidity. To maximize the pinning
strength, however, additional pining centers should be distributed
uniaxially along the same chain, rather than randomly.

Let us now study the range of SC order existing in the vortex state at
finite temperature. This can be done by considering the size dependence of
the structure factor \cite{kat93}:

\[
S\left( \overrightarrow{G}\right) =\frac{1}{N}\left\langle \left| I\left( 
\overrightarrow{G}\right) \right| ^{2}\right\rangle 
\]
where

\[
I\left( \overrightarrow{G}\right) =\int d^{2}r\left| \psi \left( 
\overrightarrow{r}\right) \right| ^{2}e^{i\left( \overrightarrow{G}\cdot 
\overrightarrow{r}\right) } 
\]
and $\overrightarrow{G}$ is a reciprocal lattice vector of the Abrikosov
lattice with $G_{x}=\frac{2\pi \nu }{a_{x}}$ , $G_{y}=\frac{2\pi m}{b_{y}}%
-2\nu b_{x}$ ; $b_{y}=\frac{\pi }{a_{x}}$ , $b_{x}=\left\langle \chi
_{n}\right\rangle a_{x}/2\pi $ , and $\nu $ , $m$ \ integers. \ \ At zero
temperature the long range order is reflected by the Bragg peaks with $%
S\left( \overrightarrow{G}\right) \propto N$ . At finite temperature: 
\begin{eqnarray}
S\left( \overrightarrow{q}\right) &=&\frac{\pi a_{x}^{2}}{2}%
e^{-q^{2}/4}\times  \label{strfc} \\
&&\sum\limits_{n,n^{\prime },\nu }\delta _{q_{x},\frac{2\pi }{a_{x}}\nu
}C_{4}\left( n^{\prime }+\nu ,n,n^{\prime },n+\nu \right) e^{-i\frac{\pi }{%
a_{x}}\left( n-n^{\prime }\right) q_{y}}  \nonumber
\end{eqnarray}
where at $\tau \gg 1$ 
\begin{eqnarray}
C_{4}\left( n_{1},n_{2},n_{3},n_{4}\right) &\equiv &\left\langle e^{i\left(
\varphi _{f}\left( n_{1}\right) +\varphi _{f}\left( n_{2}\right) -\varphi
_{f}\left( n_{3}\right) -\varphi _{f}\left( n_{4}\right) \right)
}\right\rangle  \nonumber \\
&=&\exp \left[ -\frac{s^{2}\left( p-s/3+1/3\right) }{2\tau }+i\left\langle
\chi _{n}\right\rangle \nu \left( n-n^{\prime }\right) \right]
\,\,\,\,\,\,\,\,\nu \geq 0  \label{4vcf}
\end{eqnarray}
with $s=\min \left( \nu ,\left| n^{\prime }-n\right| \right) $, $p=\max
\left( \nu ,\left| n^{\prime }-n\right| \right) $ ,$n_{1}=n^{\prime }+\nu $
, $n_{2}=n$, $n_{3}=n^{\prime }$, $n_{4}=n+\nu $. Note that $C_{4}$ is the
four-chain phase correlation function appearing in the quartic term of the
GL free energy.

Now, since $\left\langle \chi _{n}\right\rangle \equiv \overline{\chi }$ is
independent of $n$ , the sum over $n$ yields a factor $\sqrt{N}$ , and Eq.(%
\ref{strfc}) can be rewritten as:

\begin{eqnarray}
S\left( \overrightarrow{G}\right) &=&\sqrt{N}\frac{\pi a_{x}^{2}}{2}%
e^{-G^{2}/4}\times \left\{ \sum\limits_{\left| l\right| \leq \nu }\exp \left[
il\left( \nu \overline{\chi }-b_{y}G_{y}\right) -\left| l\right| ^{2}\left(
\nu -\frac{1}{3}\left| l\right| \right) /2\tau \right] \right.  \nonumber \\
&&\left. +\sum\limits_{\left| l\right| >\nu }\exp \left[ il\left( \nu 
\overline{\chi }-b_{y}G_{y}\right) -\nu ^{2}\left( \left| l\right| -\frac{1}{%
3}\nu \right) /2\tau \right] \right\}  \label{stf}
\end{eqnarray}

This expression reflects the extreme anisotropy characterizing our Bragg
chains model: Along the reciprocal lattice axis $G_{x}=0$ (i.e. for $\nu =0$
in Eq.(\ref{stf}) one finds perfect LRO, since $S\left( G_{x}=0,G_{y}\right)
\sim N$. For any $G_{x}\neq 0$ , however, the corresponding Bragg peaks
reflect only the 1D LRO within the real lattice chains, i.e. $S\left(
G_{x}\neq 0,G_{y}\right) \sim N^{1/2}$. This feature is due to the finite
range of the off-diagonal phase correlation function ( i.e. to the absence
of off-diagonal LRO ) at finite temperature.

It should be stressed here that in the triangular Abrikosov lattice there
are three equivalent ways to select the principal axes. Since the reciprocal
lattice points with $G_{x}=0$ depend on our concrete choice of the
coordinate system one may expect that by averaging over all three equivalent
orientations, the size dependence of the structure factor will be isotropic
, satisfying $S\sim N^{\sigma }$ with $1/2<\sigma <1$.

This result is consistent with the quasi LRO obtained by Kato et al. \cite
{kat93}, which is reminiscent of the
Kosterlitz-Thouless-Halperin-Nelson-Young (\cite{kthny}) theory of 2D
melting , according to which $S\sim N^{\sigma }$ with $\sigma \lesssim 5/6$.

\section{Conclusion}

In this paper we have studied the melting of the SC vortex lattice in 2D at
high magnetic fields and low temperatures (i.e. in the LLL approximation) by
using a new approximate analytical approach. Our results basically agree
with the state of the art Monte Carlo simulations. The simple analytical
approach used enables us to draw a clear picture of the melting process: The
skeleton of this picture consists of the principal crystallographic axes in
the triangular Abrikosov lattice, along which families of almost rigid Bragg
chains slide nearly freely at low temperatures due to thermal fluctuations.
Similar motions along crystal axes with higher Miller indices cost
significantly more energy and are therefore quenched at low temperatures.

The melting of the lattice occurs essentially when these fluctuations
overcome the weak attractive interaction between chains. This interaction is
not the same for the two principal axes; it is stronger for the more closely
packed family of chains.

Thus, the fluctuations destroy the order within the more loosely packed
family of chains at lower temperature. \ Consequently the configuration
based on the more closely packed family is the more stable one at low
temperatures (i.e. below $T_{m}$ ). However, at higher temperatures ( i.e.
above $T_{m}$ ) when the order in the closely packed family is also
destroyed, the configuration based on the loosely packed family of chains
becomes the more stable one , since the remaining interaction between chains
at these temperatures is repulsive and weaker for the loosely packed family.

The first order transition at $T_{m}$ is therefore a discontinuous
transformation between two different configurations of chains. The low
temperature configuration, which is characterized by small fluctuations
about the mean positions of vortices, forming an ideal Abrikosov triangular
lattice (i.e. with angle $\Theta /2=30^{\circ }$ between the principal
axes). In the high temperature configuration the phase fluctuations are
significantly larger than in the low temperature one, whereas the mean
positions of the vortices are still forming an exactly regular lattice, but
now with $\Theta \approx 75^{\circ }$.

A correlated cluster of chains nucleates only around a pinned chain. Since
according to our model, the pinning force of a whole chain can be strengthen
dramatically by distributing pinning centers uniaxially along this chain, it
may suggest a very efficient way of generating pinning defects in quasi 2D
SC. This pinning mechanism may be tested experimentally by producing
columnar defects (\cite{yesh93}) along the conducting planes in quasi 2D SC.

We would like to thank I.D. Vagner and P. Wyder for stimulating discussions
and Z. Tesanovic for helpful comments. This research was supported by THE
ISRAEL SCIENCE FOUNDATION founded by The Academy of Sciences and Humanities,
and by the fund for the promotion of research at the Technion.

\newpage %
%
{}

{\huge Figure captions:}

\vspace{1 in}{}

Fig.1 The two families of Bragg chains in the triangular lattice along the
principal axes $x$ and $x^{\prime }$. The parameters $a_{x}$ and $%
a_{x^{\prime }}$ are the periods of the order parameter modulus along these
axes respectively, while $\pi /a_{x}$ and $\pi /a_{x^{\prime }}$ are the
respective distances between chains.

Fig.2 Dependence of the Abrikosov parameter, $\beta _{a}$, on $z=\left( \pi
/a_{x}\right) ^{2}$. The two minima at $z_{1}=\pi /2\sqrt{3}$ and $z_{2}=%
\sqrt{3}\pi /2$ correspond to the triangular Abrikosov lattice, $\beta
_{a}=\beta _{A}$, with different choice of the Bragg chains direction. The
maximum at $z=\pi /2$ corresponds to the square lattice.

Fig.3 Free energy of fluctuating Bragg chains (normalized by the mean field
free energy), $-\beta _{A}/\beta _{fl}$, as a function of the reduced
temperature, $t=-\frac{2\pi \alpha ^{2}}{\beta k_{B}T}$. Solid and dash
lines correspond to Bragg chains along the $x$ and $x^{\prime }$ directions
respectively. The intersection point at $t=t_{m}\approx -16$ determines the
phase transition.

Fig.4 Dependence of the shear modulus, $\mu $, on the reduced temperature, $%
t $. The jump of $\mu $ at $t=t_{m}$ reflects the 'melting' transition.


\begin{references}
\bibitem{ftn1}  For example, by adding $N_{0}$ degrees of freedom to each
unit cell , coresponding to the wavenumbers $q_{n,m}=\frac{2\pi }{L_{x}}%
\left( \sqrt{N}n+m\right) $ $\left( m=-N_{0}/2+1,.....,N_{0}/2\right) $, and
taking the coefficients $c_{n,m}$ with equal amplitudes , the superfluid
density can be shown to be localized along the $x$ direction with size $\sim
L_{x}/N_{0}$. For such a state we find $\beta _{a}\sim N_{0}$ , and the
corresponding SC condensation energy is by the factor $1/N_{0}$ smaller than
the mean field condensation energy. Also, by allowing the positions of $%
c_{n} $ to deviate from the equlibrium periodic array $q_{n}=2\pi n/a_{x}$,
the mean values of $\left| c_{n}\right| ^{2}$ are no longer independent of $%
n $, and the corresponding fluctuations cause variations in the free energy
functional of lower orders than $\lambda ^{2}$. Consequently, this type of
fluctuations are also significantly more energetic than the phase
fluctuations considered in the present paper.
\end{references}
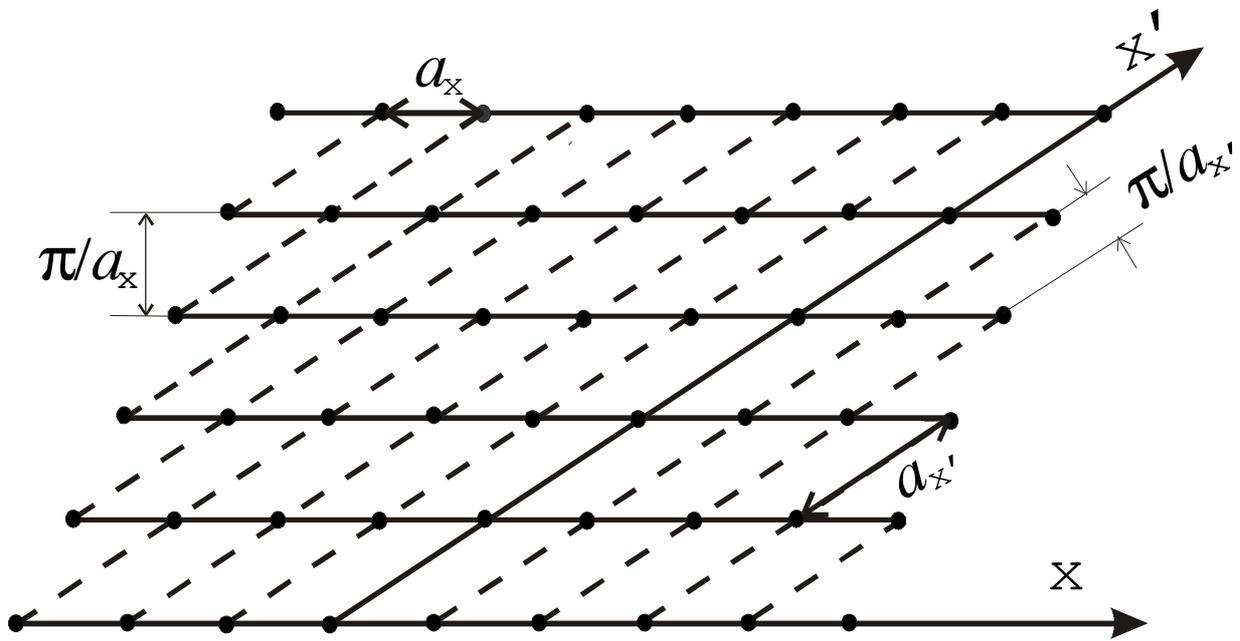
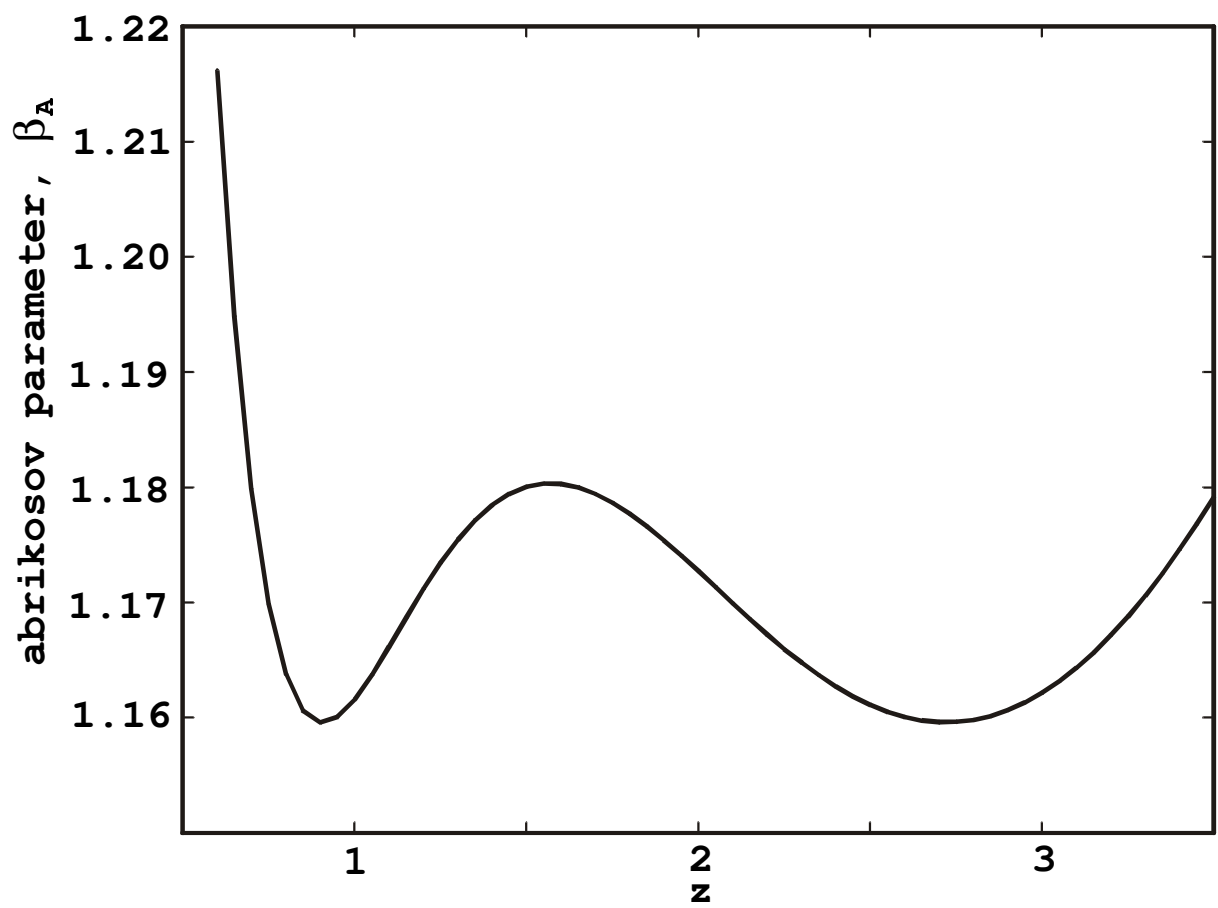
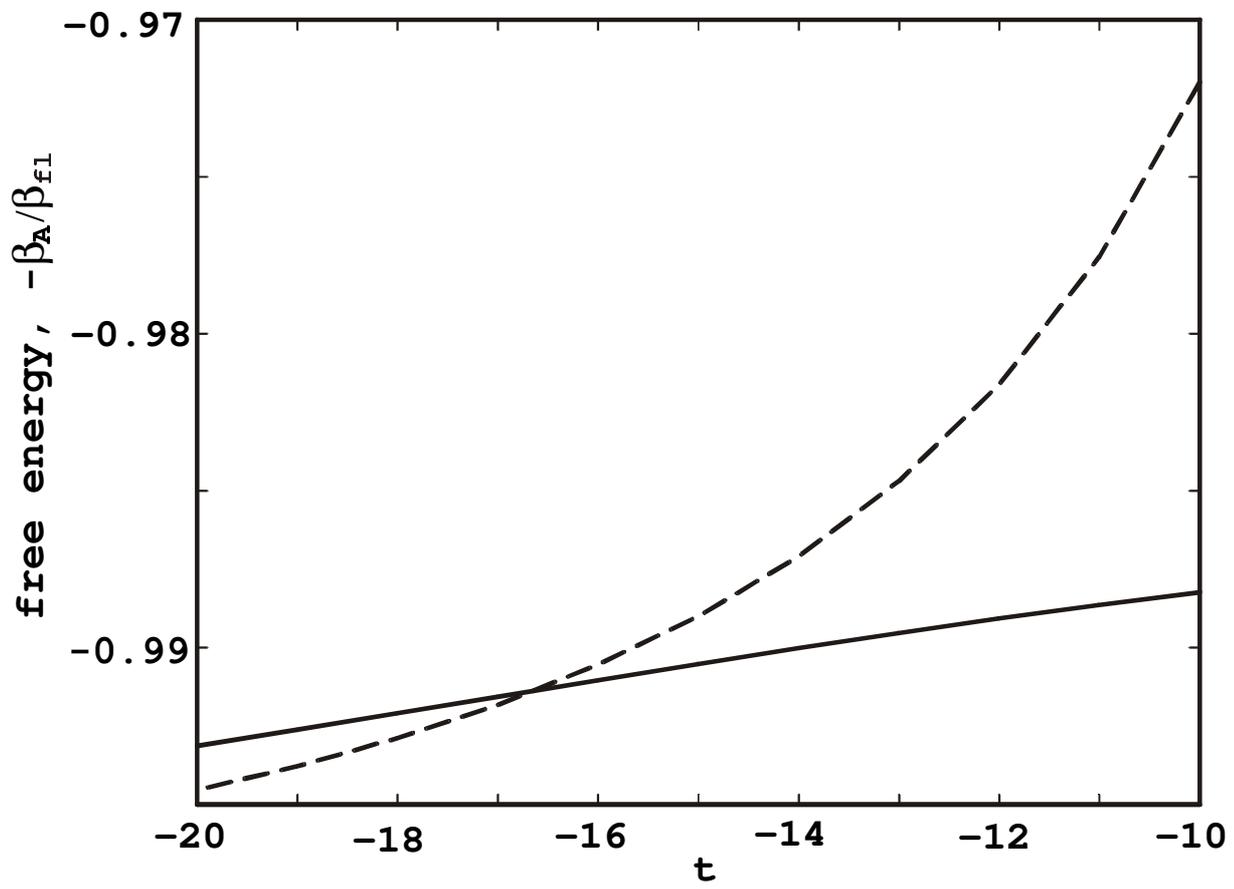
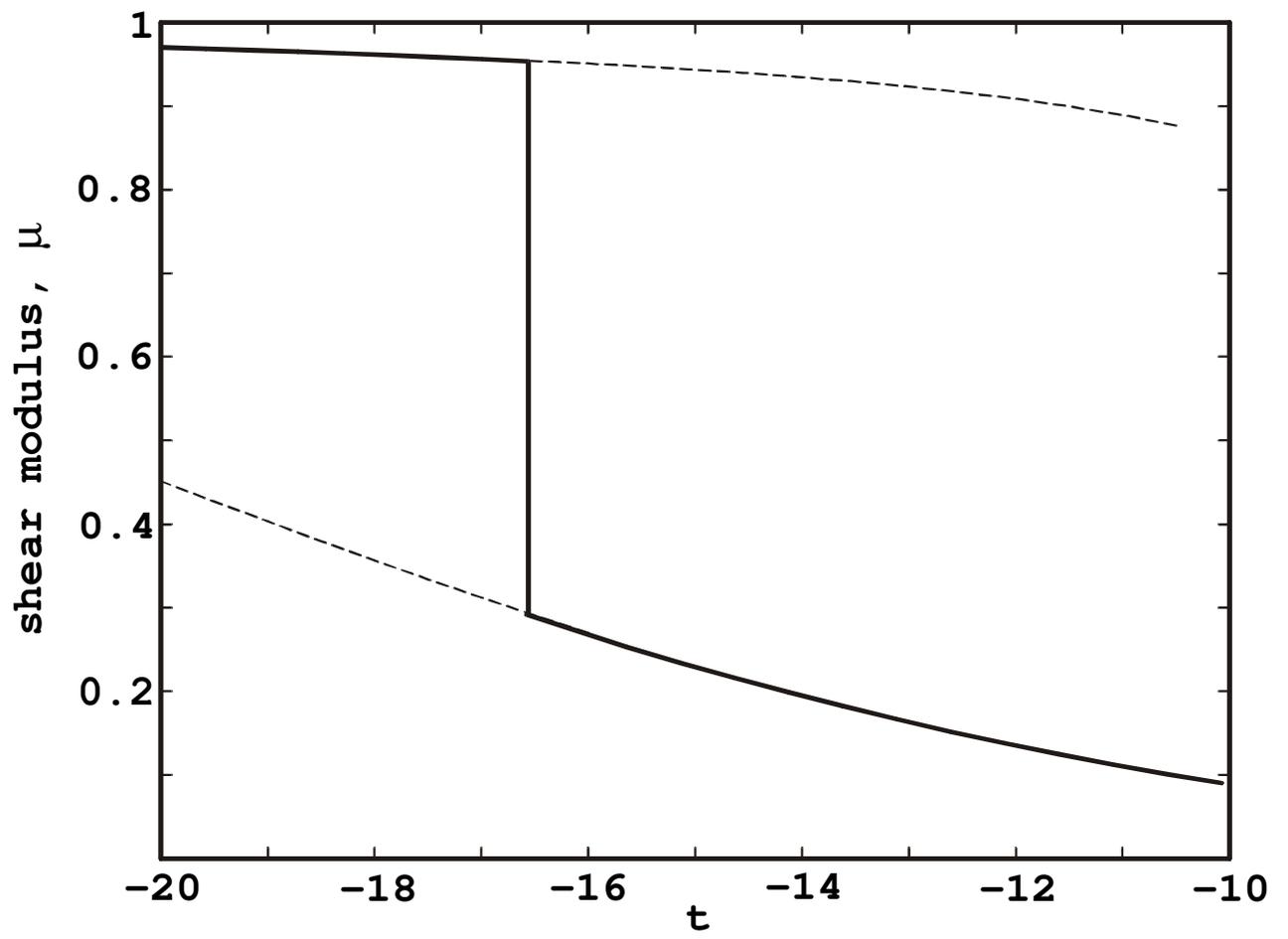
\end{document}